\documentstyle[12pt,aaspp4]{article}

\received{}
\revised{}
\accepted{}
\cpright{}{1999}

\journalid{}{}
\articleid{}{}
\paperid{}

\cpright{}{}
\ccc{}
\begin{document}

\newcommand{\etal}{et al.~}
\newcommand{\bperp}{$B_\perp$}
\newcommand{\blos}{$B_{||}$}
\newcommand{\degr}{$^\circ$}

\title{Magnetic Fields in Star-Forming Molecular Clouds I. The First
Polarimetry of OMC-3 in Orion A}

\author{Brenda C.~Matthews\altaffilmark{1}}
\author{Christine D.~Wilson\altaffilmark{1}}

\altaffiltext{1}{Department of Physics \& Astronomy, McMaster University, Hamilton,
Ontario, Canada, L8S 4M1}

\begin{abstract}

The first polarimetric images of the OMC-3 region of the Orion A
filamentary molecular cloud are presented.  Using the new imaging
polarimeter on SCUBA at the James Clerk Maxwell Telescope, we have
detected polarized thermal emission at 850 \micron~from dust along a
6\arcmin~length of the dense filament.  The polarization pattern is
highly ordered and is aligned with the long axis of the filament
throughout most of the region, diverging only near the southern
boundary by $30-50$\degr.  If the polarization arises from thermal
emission of dust grains aligned via either paramagnetic inclusions or
radiative torques, this configuration indicates a plane-of-sky
magnetic field which is normal to the filament along most of its
length.  The mean percentage polarization is 4.2\% with a 1$\sigma$
dispersion of 1\%.  This region is part of the integral-shaped
filament, and active star formation is ongoing along its length, with
only two of nine dust condensations in our field lacking evidence of
outflow activity.  The outflow directions do not appear to be
consistently correlated with the direction of the plane-of-sky field
or the filament structure itself.  Depolarization toward the filament
center, previously detected in many other star-forming cores and
protostars, is also evident in our data.

\end{abstract}


\section{Introduction}

It is now well established that magnetic fields play a significant
role in the evolution of molecular clouds and their associated star
formation (see Heiles \etal 1993 and references therein).  At a
distance of 500 pc, the Orion complex is the closest star forming
region that is undergoing massive star formation, and as such, has
been the object of intense study.  Many studies have focused on the
region known as OMC-1, a massive cloud core which lies behind the
Orion Nebula (M42).  This core is embedded in the integral-shaped
filament, identified in the $^{13}$CO $J=1-0$ transition by Bally
\etal (1987) and most recently mapped at 850 and 450 \micron~by
Johnstone \& Bally (1999).  The OMC-3 region lies at the northern tip
of the filament (Bally \etal 1987), near the HII region NGC 1977 (see
Kutner, Evans \& Tucker 1976).  Molecular studies reveal the dust
temperatures to be considerably cooler (T $\sim$ 20-25 K) in the OMC-3
region than in OMC-1 (Chini \etal 1997).

Dust condensations in OMC-3 were identified in 1.3 mm continuum by
Chini \etal (1997).  An evolutionary sequence has been suggested, with
source ages declining as one moves north along the filament.
Reipurth, Rodr\'{\i}guez \& Chini (1999) have done a VLA search at 6
cm for compact sources in the region and report no sources coincident
with the positions of MMS1 and MMS4, two of the most northern
condensations identified by Chini \etal (1997).  Additionally, there
is no outflow associated with either source (Yu, Bally \& Devine 1999;
Chini \etal 1997; Castets \& Langer 1995).  These results suggest that
these objects may in fact be in a pre-collapse phase, since outflows
are known to be associated with the earliest phases of collapse (Shu,
Adams \& Lizano 1987).

Goodman \etal (1995) illustrated that measurements of polarization of
background starlight in the optical and near-infrared are not
effective tracers of magnetic field structure in dense molecular gas
due to poor alignment and/or amorphous dust grain structure.  At
submillimeter and millimeter wavelengths, aligned, rotating grains
produce polarized thermal emission.  Draine \& Weingartner (1996,
1997) have shown that radiative torques are highly effective at
aligning grains of specific sizes ($\sim$0.2 $\mu$m).  Lazarian,
Goodman \& Myers (1997) suggest that the effects of radiative torques
could be comparable to paramagnetic inclusions (Purcell 1975, 1979;
Spitzer \& McGlynn 1979) at aligning grains in the star-forming
interstellar medium.  Both these mechanisms result in grains aligned
with their long axes perpendicular to the magnetic field.

In this paper, we present 850 \micron~imaging polarimetry of OMC-3.
These data represent the first polarimetry of this active star-forming
region and are also some of the first obtained with the new imaging
polarimeter at the James Clerk Maxwell Telescope (JCMT).  This is the
first publication from a project designed to study magnetic field
structure in a variety of molecular clouds in different phases of star
formation.  In $\S$2, we describe the polarimeter and the observing
and reduction techniques; in $\S$3, a mosaic of OMC-3 is presented and
discussed; $\S$4 summarizes the results thus far.

\section{Observations and Data Reduction}
\label{obs}

The observations were taken from 1998 September 5 to 7 using the new
imaging polarimeter on SCUBA (Submillimeter Common User Bolometric
Array) at the JCMT.  These nights were stable, with $\tau$(225 GHz)
ranging from 0.05 to 0.07 during the period of observations.
Calibration of the polarizer was performed on 1998 September 5 using
the Crab Nebula, for which percentage polarization $p = 19.3\pm
4$\% and position angle $\theta = 155\pm$ 5\degr~were measured.

The polarimeter consists of a rotating quartz half-waveplate and a
fixed analyzer which are used to measure linear polarization of
thermal emission.  The waveplate introduces a phase-lag of one half
wavelength between the plane of polarization along the `fast axis' and
the orthogonal plane.  Rotation of the plate changes the angle between
the fast axis and the plane of the incoming source polarization, so a
varying component of the polarized emission is retarded.  The analyzer
is a photo-lithographically etched grid of 6 $\mu$m spacing which
transmits only one plane of polarization (by absorbing photons with an
{\bf E}-component parallel to the wires) to SCUBA.  The detector thus
sees a modulated signal as the waveplate rotates.  The variations are
used to deduce the percentage polarization and polarization position
angle of the source.

Seven pointing centers were observed along the OMC-3 filament, four
the first night and only three (slightly shifted in position) on
subsequent nights to provide better coverage of low signal-to-noise
regions.  Data for each pointing center were co-added to create seven
maps of $I$, $Q=q/I$ and $U=u/I$, where $I$, $q$ and $u$ are Stokes'
parameters.  Each polarization cycle consists of 16 integrations at
22.5\degr~rotation intervals (i.e.~4 independent measurements at each
angle) in 8 minutes integration time.  The brightest source, MMS6, was
observed for only 6 cycles, or 48 minutes integration time.
Observations centered on MMS1 and MMS4 required $\sim$3 hours
integration time.  The sensitivity of the SCUBA detector yielded
polarization measurements on sources of 0.5 Jy beam$^{-1}$ at the
6$\sigma$ level for 4\% polarized flux.

For each polarization cycle map, the standard preliminary reduction
was done.  This included subtraction of sky levels as well as
instrumental polarization (IP) for each bolometer.  Polarizations less
than 1\% are unreliable, since the IP has been found to vary by $\pm
0.5$\% for about 30\% of the bolometers.  The average IP for all
off-axis bolometers is $0.88\pm 0.06$\% @ $166 \pm$ 2\degr~while the
IP of the central bolometer is $1.08 \pm 0.10$\% @ $158 \pm$ 3\degr.
After correction for source rotation across the array, the Stokes'
parameters were calculated by comparing measurements offset by
45\degr~in waveplate rotation (90\degr~on the sky).  The $I$, $Q$ and
$U$ maps for each pointing center were averaged, and standard
deviations were derived by comparing the individual data sets.  The
maps were then binned spatially by a factor of two to yield
6\arcsec~(approximately half-beamwidth) sampling.

At this point, it was necessary to diverge from the standard reduction
technique to make mosaics of the $I$, $Q$ and $U$ maps before
calculating the percentage polarization $p$ and position angle
$\theta$.  Mosaicing was done using the MAKEMOS tool in Starlink's
CCDPACK (a UK based software package).  Variances were used to weight
the overlapping data values, and variances were also generated for the
map.  The calculation of $p$ (and its uncertainty) is given by:
\begin{equation} 
p = \sqrt{Q^2 + U^2}; \ \ \  dp= p^{-1}\sqrt{[dQ^2Q^2 + dU^2U^2]}.
\end{equation}

\noindent A bias exists which tends to increase the $p$ value, even
when $Q$ and $U$ are consistent with $p=0$ because $p$ is forced to be
positive.  The polarization percentages were debiased according to the
expression:
\begin{equation}
p_{db}= \sqrt{p^2-dp^2}.
\end{equation}

\noindent The position angle can then be calculated by the following
relations:
\begin{equation}
\theta = 0.5 \arctan(U/Q); \ \ \ d\theta = 28.6^\circ / \sigma_p
\label{dpa}
\end{equation}

\noindent where $\sigma_p$ is the ratio $p_{db}/dp$.  

The $p_{db}$ values were then thresholded such that $p_{db} \le$ 100\% and 
$\sigma_p \ge 2.5$.  The position angles can of course take on any value,
but we note that offsets of 180\degr~cannot be distinguished in linear
polarization.

\section{Results}

\subsection{The Polarization Pattern}

Figure \ref{THEMAP} shows the OMC-3 filament in 850 \micron~continuum
(colored greyscale) overlaid with polarization vectors.  Only vectors
up to 20\% are plotted on the figure; 27 vectors in total were
omitted.  The blue contours indicate where $\sigma_p=6$ and $\sigma_p
= 10$.  By equation (\ref{dpa}), the vectors enclosed by these
contours have $d\theta \le 5^\circ$ and $d\theta \le 3^\circ$,
respectively.  No vectors have $\sigma_p < 2.5$, so no plotted vector
has $d\theta > 12^\circ$.

The polarization vectors are well-ordered along the filament with the
best alignment in the northern region between the sources MMS1 and
MMS6 (as identified by Chini \etal 1997, see Figure \ref{THEMAP}).
Since the vectors show a high degree of ordering, it is tempting to
think that they are uniform across the field.  However, extraction of
data subsets centered on each of the four ``regions'' indicated in
Figure \ref{THEMAP} illustrates that this is not the case.  The
distribution of $\theta$ in each region is fit by either 1 or 2
Gaussians by minimizing chi-squared.  The position angle (mean and
dispersion) of each of these Gaussians is noted on the figure, as well
as the reduced chi-squared of each fit.

The dispersions in $\theta$ are relatively narrow, when compared with
the sample presented in Myers \& Goodman (1991) for 26 dark cloud
regions.  If a single gaussian is fit to every subregions'
distribution (regardless of the goodness-of-fit), it is found that
Regions A through D have dispersions of 8, 8, 9 and 12$^\circ$
respectively.  Such low dispersions were identified only for dark
clouds without clusters, i.e. less than 15 associated stars in 1
pc$^2$ (Myers \& Goodman 1991), yet OMC-3 has condensation density of
135 pc$^{-2}$ (taking 9 sources in a $6^\prime \times
30^{\prime\prime}$ area).  According to the Myers \& Goodman model of
uniform and non-uniform field components, this result implies either a
low ratio of non-uniform to plane-of-sky uniform components of the
magnetic field, or a low number of magnetic field correlation lengths
along the line of sight.  To implement their analysis fully will
require measurement of the line-of-sight component of the magnetic
field toward several positions in OMC-3.

It is interesting to compare the changes in the position angle of the
filament on the sky with the change in orientation of the polarization
vectors.  The filament's orientation can be traced easily due to the
positions of the condensations themselves, which without exception are
embedded within it.  Measuring an angle E of N, three main segments of
OMC-3 can be distinguished.  From MMS1 to MMS6, the filament is at an
angle of $\sim$130\degr~($-50$\degr); this area is covered by Regions
A and B as denoted on Figure \ref{THEMAP}. These histograms reveal
that the peaks fit to these distributions agree with the position
angle of the filament to within 10\degr.  The situation is similar for
Region C, which contains MMS7, the only IRAS source in OMC-3
(05329$-$0505).  The angle of the filament steepens from MMS6 to MMS7
to $\sim$160\degr~($-20$\degr).  The distribution of $\theta$ exhibits
two peaks in position angle.  The strongest peak is in fact at $-19
\pm 7$\degr~which indicates excellent alignment with the filament.
(The uncertainty represents the 1$\sigma$ dispersion in the
distribution.)  However, a second peak exists at $-36 \pm 5$\degr.
Finally, from MMS7 through MMS9, the filament aligns north-south
(position angle 0\degr).  None of the polarization vectors exhibit
such an angle; instead, the distribution is double peaked at $-33 \pm
5$\degr~and $-47\pm 15$\degr.  Since the maximum uncertainty
associated with any value of $\theta$ is $12^\circ$, Region D is the
only one where no alignment exists between the filament and the
polarization pattern.

In short, the polarization, and hence inferred field direction (in the
plane of the sky), \bperp, changes as one moves along OMC-3.
\bperp~is predominantly perpendicular to the filament along most of
its length, diverging by $30-50$\degr~from the filament only in the
southernmost part of OMC-3.  These results are roughly consistent with
the work of Schleuning (1998), which also established
\bperp~perpendicular to the filament direction in OMC-1.  The field
direction does not appear to be affected by the presence of the dust
condensations, but rather is aligned with the structure of the
filament itself.  At a resolution of 15\arcsec~(7500 A.U.), these data
simply may not have enough resolution to detect the details of fields
associated with the starless cores or protostellar envelopes and their
associated outflows.

\subsection{The Influence of Outflows?}

Many previous works have noted that the inferred \bperp~field
direction from long wavelength polarization is oriented either
parallel (e.g.~IRAS 16293, Tamura \etal 1993; NGC 1333, Minchin,
Sandell \& Murray 1995; Tamura, Hough \& Hayashi 1995) or
perpendicular (e.g.~VLA 1623, Holland \etal 1996) to the observed
protostellar outflow direction. The relative orientations of the
outflows in OMC-3 are illustrated by the green lines on Figure
\ref{THEMAP}, as measured by Chini \etal (1997) and Yu, Bally \&
Devine (1997) using $^{12}$CO $J=2-1$ and H$_2$ shocks, respectively,
to identify outflow signatures.  With the exception of MMS6, the
outflows are aligned E-W.  Hence, in Regions A-C, the outflows are
perpendicular neither to the filament nor the \bperp-field.  In Region
D, they are aligned perpendicular to the filament, but are offset from
\bperp~by $\sim 30-50^\circ$; however, it is possible that we may be
detecting a superposition of the magnetic fields of the filament and
the outflow(s).  Reipurth, Rodr\'{\i}guez \& Chini (1999) suggest that
MMS9 is in fact the driving source for the most powerful outflow in
the OMC-3 region.  If the evolutionary sequence proposed by Chini
\etal (1997) is correct, then MMS9's outflow may have had sufficient
time to alter the magnetic field in its vicinity.

These results suggest that the field of the filament alone does not
determine the outflow direction.  Thus, there must be other relevant
factors which determine the stucture of a protostellar system.  Were
these data interpreted as a uniform field along which material had
collapsed, one would naively expect protostellar disks to be aligned
parallel to the filament, and outflows perpendicular to it but aligned
with the ambient field; however, this is not observed.

\subsection{Percentage Polarization}

The distribution of $p_{db}$ along the filament exhibits a mean value
of 4.2\%, with a 1$\sigma$ dispersion of 1\%.  Values up to 100\% are
allowed, but only those $< 20$\% (i.e.~all but 27) are plotted on
Figure \ref{THEMAP} since larger values are unlikely to be physical.
The rms $dp$ and $\sigma_p$ values of unplotted data are 15\% and 2.9,
compared to 2.5\% and 7.3 for all values of $p$ with $\sigma_p > 2.5$.
Polarizations up to 11.9\% have been detected with $\sigma_p \sim 7$.

Several authors have discussed whether observations of decreased
polarization percentage toward regions of higher flux are due to
changes in physical conditions or averaging of small scale variations
in a large beam.  Figure \ref{depol} shows the percentage polarization
along a cut perpendicular to the filament through MMS4 (indicated by
the red cross in Figure \ref{THEMAP}, which has the highest $\sigma_p$
in our data set.  Although the uncertainties are increasing toward the
edge of the filament, the data show a clear trend of decreasing
polarization percentage toward the filament's center.  These changes
could be due to changes in either grain properties or field strength,
but they have also been suggested as an observable signature of
helical fields (Fiege \& Pudritz 1999).  The interpretation of this
depolarization effect will be discussed more fully in a forthcoming
paper.

\section{Summary}

Submillimeter wavelength polarimetry with the JCMT has revealed a
highly ordered polarization pattern along the filament known as OMC-3.
These data indicate that \bperp is perpendicular to the filament along
most of its length, diverging only in the most southern regions by
between $30-50$\degr.  The outflows which have been observed in OMC-3
are aligned with neither the ambient field nor the filament in any
consistent way.  The field of the filament is thus unlikely to be the
dominant factor in determining the configuration of the protostellar
systems embedded within it.  The mean percentage polarization is
4.2\%, with a 1$\sigma$ dispersion of 1\%.  Values as high as 11.9\%
have been measured with $\sigma_p \sim 7$.  A depolarization effect is
measured toward the denser parts of the filament.

\bigskip

The authors would like to thank J. Greaves, T. Jenness, G.
Moriarty-Schieven and A. Chrysostomou at the JCMT for their assistance
with problems both large and small during and especially after
observing, and the referee for an insightful and thorough review.  BCM
would like to thank J. Fiege for many constructive conversations and
MATLAB expertise.  The research of BCM and CDW is supported through
grants from the Natural Sciences and Engineering Research Council of
Canada.  The JCMT is operated by the Joint Astronomy Centre on behalf
of the Particle Physics and Astronomy Research Council of the UK, the
Netherlands Organization for Scientific Research, and the National
Research Council of Canada.

\clearpage

\figcaption[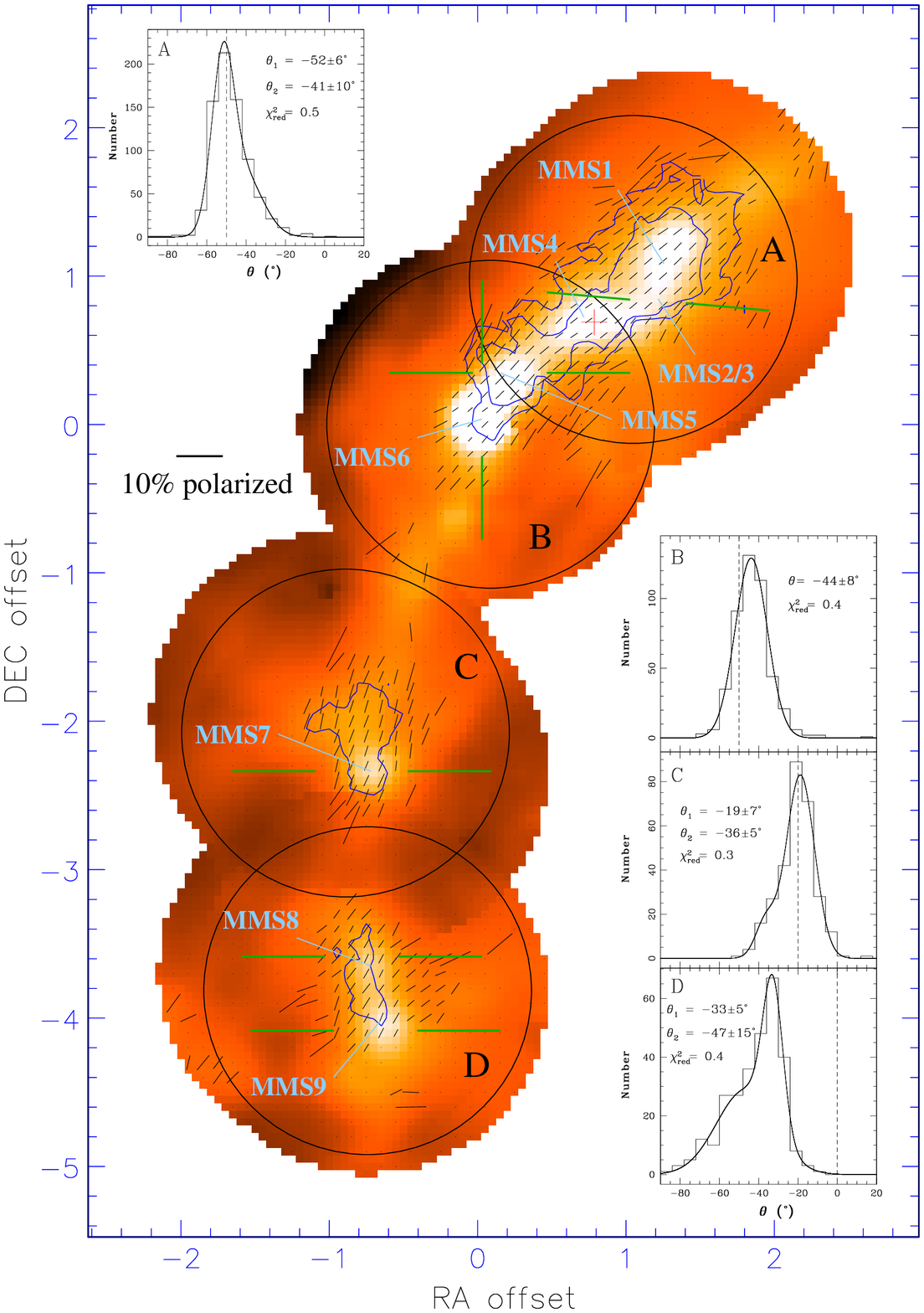]{850 \micron~polarized emission along OMC-3.
The coordinates are arcminute offsets from the position of MMS6,
$\alpha_{1950} = 5^{\rm h} 32^{\rm m} 55.6^{\rm s}$ and $\delta_{1950}
= -5^\circ 03\arcmin ~25\arcsec$.  Color greyscale indicates
variations in uncalibrated I, with a range of $-2$ to 3 $\sigma$.
Polarization vectors (to a maximum of 20\%) are overlain.  All vectors
shown have $\sigma_p > 2.5$.  Blue contours indicate bounds of
$\sigma_p =6$ and $\sigma_p=10$ respectively.  These bounds also
indicate where $d\theta < 5^\circ$ and $d\theta < 3^\circ$
respectively.  The condensations of Chini \etal (1997) are labelled
MMS1 through MMS9, and outflow orientations are shown in green.  The
red cross shows the center of the cut shown in Figure \ref{depol}.
Subregions, labelled A through D, are identified, and the
distributions of position angle for each are shown in the histograms.
Position angles are grouped to 6$^\circ$ bin widths.  To test the form
of the distributions, binning was also done with widths of 3$^\circ$
and 12$^\circ$.  The only cases in which the distributions were not
consistent were Regions A and D with a 12\degr~bin width.  In these
cases, the double Gaussian profile did not produce a fit to the
data. Nevertheless, all single Gaussian fits for these two regions
produced reduced chi-squared values $> 1$; hence, single Gaussians are
poor fits to these distributions.  The parameters of the fits are
noted on each figure, as well as the reduced chi-square values.  The
dashed lines represent the mean position angle of the {\it filament}
in each region.
\label{THEMAP}}

\figcaption[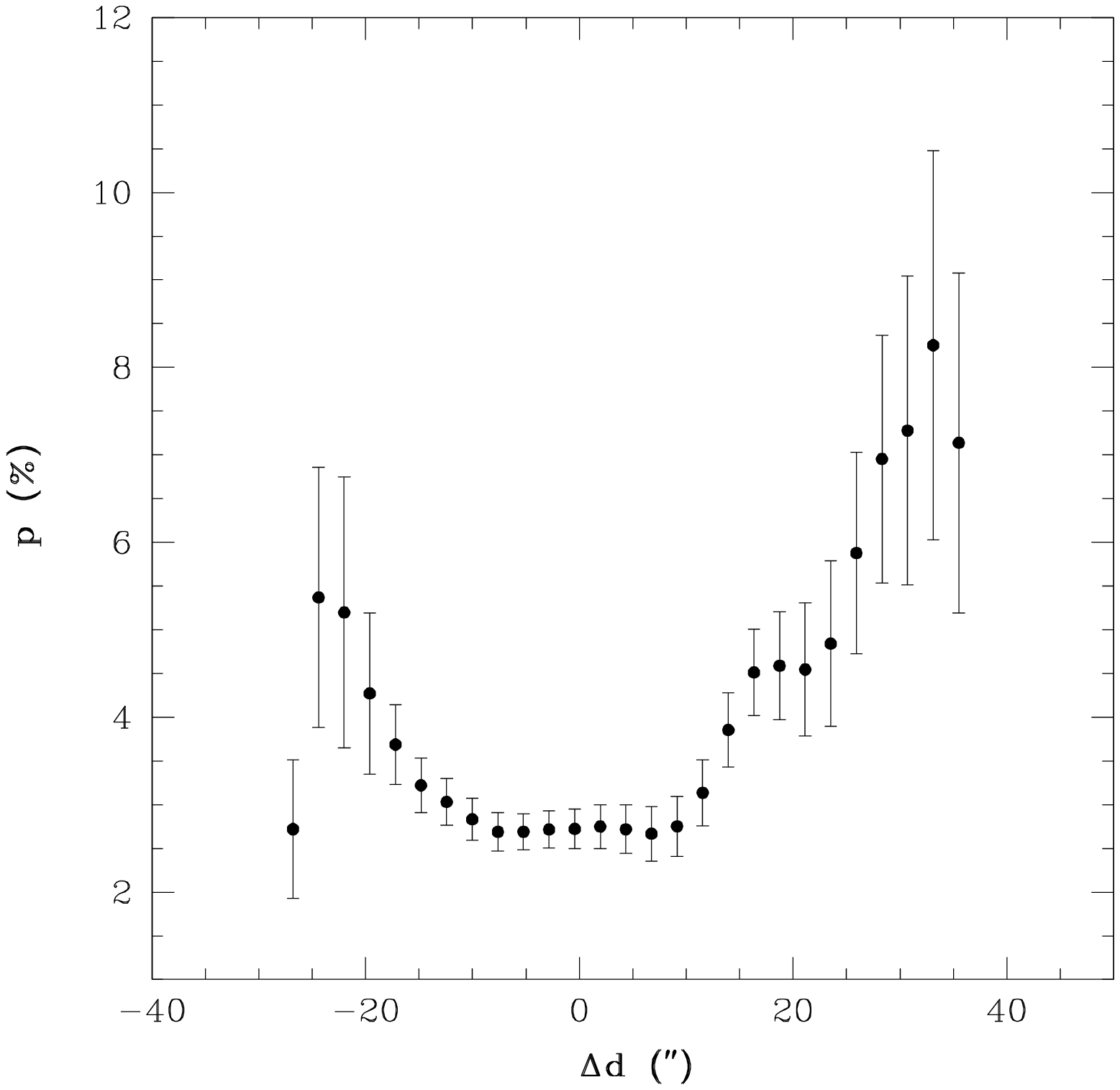]{Variation in $p_{db}$ along a line
perpendicular to the filament.  These data are centered on
$\alpha_{1950} = 5^{\rm h} 32^{\rm m} 52.5^{\rm s}$ and $\delta_{1950}
= -5^\circ 02\arcmin ~46\arcsec$, the position of the source MMS4
identified by Chini \etal (1997) and identified by the red cross in
Figure \ref{THEMAP}.  The cut is centered on MMS4, and $\Delta d$ is
the offset perpendicular to the filament.  The plot clearly shows a
reduction in polarization toward the center of the filament.  The
error bars represent the uncertainty associated with each interpolated
data point.  The depolarization effect cannot be accounted for merely
by the increasing noise in regions of low signal.  Only data points
with $\sigma_p > 2.5$ were used.
\label{depol}}

\end{document}